\documentclass[aps,pre,twocolumn,showpacs,floatfix]{revtex4} 
\usepackage{amsmath,amssymb,isolatin1,graphicx,times}

\newcommand{\be}{\begin{equation}}
\newcommand{\ee}{\end{equation}}
\newcommand{\bea}{\begin{eqnarray}}
\newcommand{\eea}{\end{eqnarray}}

\newcommand{\kommentar}[1]{}

\begin{document}
 
\title{Asymmetries in symmetric quantum walks on two-dimensional networks}
\author{Oliver M{\"u}lken}
\email{oliver.muelken@physik.uni-freiburg.de}
\author{Antonio Volta}
\email{antonio.volta@physik.uni-freiburg.de}
\author{Alexander Blumen}
\email{blumen@physik.uni-freiburg.de}
\affiliation{
Theoretische Polymerphysik, Universit\"at Freiburg,
Hermann-Herder-Straße 3, D-79104 Freiburg, Germany}

\date{\today} 
\begin{abstract}
We study numerically the behavior of continuous-time quantum walks over
networks which are topologically equivalent to square lattices. On short
time scales, when placing the initial excitation at a corner of the
network, we observe a fast, directed transport through the network to the
opposite corner. This transport is not ballistic in nature, but rather
produced by quantum mechanical interference. In the long time limit,
certain walks show an asymmetric limiting probability distribution; this
feature depends on the starting site and, remarkably, on the precise size
of the network. The limiting probability distributions show patterns which
are correlated with the initial condition. This might have 
consequences for the application of continuous time quantum walk
algorithms.
\end{abstract}
\maketitle

\section{Introduction}

The study of quantum mechanical extensions of classical transport
processes has witnessed a considerable growth over the last decade, with
much attention being given to quantum information and the application of
transport processes in potential quantum computers \cite{Nielsen}. This
led to the development of several quantum algorithms specifically designed
for these quantum computers. 
Among these algorithms are the
so-called quantum walks, see for instance \cite{shenvi2003,childs2004},
which are the analog of the classical random walks
\cite{aharonov1993,farhi1998}, for an overview see \cite{kempe2003}.

Such quantum walks have been studied in different situations. For
instance, it was shown that continuous-time quantum walks can cause a
remarkable speed-up of transport through certain graphs \cite{childs2002},
a feature which is not universal, as discussed in Ref.\ \cite{mb2005a},
where it was shown that quantum transport can also become much slower than
the classical one, depending on the initial conditions. Another variant of
the quantum walk, the quantum mulitbaker map, was shown to exhibit a crossover
from classical to quantal behavior with time, where the crossover time is
given by the inverse of Planck's constant \cite{wojcik2002,wojcik2003}.

For simple structures these quantum walks are directly related to well
known problems in solid state physics. Thus, methods used in solid state
physics can easily be applied to quantum walks. For instance, walks on
one-dimensional lattice with periodic boundary conditions are readily
treated by a Bloch ansatz \cite{mb2005b,Ziman,Kittel}.

Here, we study quantum walks on two-dimensional networks. We focus on
structures topologically equivalent to square lattices.  For these we
evaluate the transition probabilities between different nodes of the
finite networks and compare these to the Bloch solutions.  As we are going
to show, these transition probabilities display odd, unexpected features.

The paper is organized as follows: In Sec.\ref{ctqw} we briefly review the
properties of continuous time quantum walks and the Bloch ansatz. The time
dependent transition probabilities between the nodes of the network are
presented in Sec.\ref{pd}.  We conclude with a summary of results in
Sec.\ref{con}.

\section{Continuous time quantum walks} \label{ctqw}

In quantum information theory, qubits on a graph are used to define the
quantum analog of a random walk. There is a discrete \cite{aharonov1993}
and a continuous-time \cite{farhi1998} version.  However, distinct from
classical physics, these two are not equivalent to each other. Here, we
will focus on continuous-time quantum walks on networks. 

\subsection{Continuous time quantum walks on graphs}

We consider two-dimensional graphs, topologically equivalent
to finite, square lattices of side length $N$. In this way the nodes of
the graph are connected in a very regular manner.  In general, to every
graph there corresponds a discrete Laplace operator, sometimes called the
adjacency or connectivity matrix ${\bf A} = (A_{\boldsymbol{ij}})$,
defined by letting the non-diagonal elements $A_{\boldsymbol{ij}}$ equal
$-1$ if nodes $\boldsymbol i$ and $\boldsymbol j$ are connected by a bond
and $0$ otherwise. The diagonal element $A_{\boldsymbol{ii}}$ is given by
the number of bonds which exit from node $\boldsymbol i$, i.e.,
$A_{\boldsymbol{ii}}$ equals the functionality $f_{\boldsymbol i}$ of the
node. Thus, in our case we have $f_{\boldsymbol i} =2$ if node
$\boldsymbol i$ is located at a corner of the square, $f_{\boldsymbol
i}=3$ if the node is located along an edge (and is not a corner node), and
$f_{\boldsymbol i}=4$ otherwise. 

Classically, continuous-time random walks (CTRWs) are described by
the master equation \cite{weiss,vankampen}
\be
\frac{d}{d t} p_{\boldsymbol{k,j}}(t) = \sum_l
T_{\boldsymbol{kl}} \ p_{\boldsymbol{l,j}}(t),
\label{mast_eq0}
\ee
where $p_{\boldsymbol{k,j}}(t)$ is the conditional probability to find the
walker at time $t$ at node $\boldsymbol k$ when starting at time $0$ at
node $\boldsymbol j$.  We assume an unbiased CTRW such that the
transmission rate $\gamma$ of all bonds are equal. Then the transfer
matrix of the walk, ${\bf T} = (T_{\boldsymbol{kj}})$, is related to the
adjacency matrix by ${\bf T} = - \gamma {\bf A}$.  Formally,
Eq.(\ref{mast_eq0}) is solved by
\be
p_{\boldsymbol{k,j}}(t) = \langle \boldsymbol k | e^{{\bf T} t} |
\boldsymbol j \rangle.
\label{cl_prob}
\ee

The quantum mechanical extension of a CTRW, the continuous-time quantum
walk (CTQW), is now defined by identifying the Hamiltonian of the system
with the (classical) transfer operator, ${\bf H} = - {\bf T}$,
\cite{farhi1998,childs2002,mb2005a}. 
However, there is no unique way of defining a CTQW. As mentioned in
\cite{childs2004}, there is some freedom in defining the diagonal elements
of the Hamiltonian for a certain graph. A requirement for the CTQW is that
${\bf H}$ define a unitary process, whereas classically the CTRW is a
probability conserving Markov processes, which requires $\sum_{\boldsymbol
k} T_{\boldsymbol{kj}} =0$.  As mentioned in \cite{childs2004}, for
regular networks, where all nodes have the same functionality, different
choices of the Hamiltonian give rise to the same quantum dynamics.
Nevertheless, in what follows we directly identify the Hamiltonian with
the transfer operator since some of the networks we consider are
non-regular. 

In the spirit of a localized orbital
picture we now introduce the states $|\boldsymbol j\rangle$ which are
localized at the nodes $\boldsymbol j$ of the graph, requiring in an
obvious notation that $\langle \boldsymbol k | \boldsymbol j \rangle =
\delta_{\boldsymbol{k,j}}$ and that $\sum_{\boldsymbol j} | \boldsymbol j
\rangle \langle \boldsymbol j | = \boldsymbol 1$ where $\boldsymbol 1$ is
the identity operator. Under these conditions the Schr\"odinger equation
(SE) reads
\be
i \frac{d}{d t} | \boldsymbol j \rangle = {\bf H} | \boldsymbol j \rangle,
\label{sgl}
\ee
where we set $\hbar=1$. Starting at time $t_0$ from the state $|
\boldsymbol j \rangle$ the system evolves as ${\bf U}(t,t_0) | \boldsymbol
j \rangle$, where ${\bf U}(t,t_0) = \exp(-i {\bf H} (t-t_0))$ is the
quantum mechanical time evolution operator. Hence, the transition
amplitude $\alpha_{\boldsymbol{k,j}}(t)$ from state $| \boldsymbol j
\rangle$ at time $0$ to state $| \boldsymbol k\rangle$ at time $t$ is
\be
\alpha_{\boldsymbol{k,j}}(t) = \langle \boldsymbol k | e^{-i {\bf H} t} |
\boldsymbol j \rangle.
\label{qm_ampl}
\ee
With Eq.(\ref{sgl}) we find that
\be
i \frac{d}{d t} \alpha_{\boldsymbol{k,j}}(t) = \sum_l H_{\boldsymbol{kl}}
\alpha_{\boldsymbol{l,j}}(t).
\label{sgl_ampl}
\ee

We note that the main difference between Eq.(\ref{cl_prob}) and
Eq.(\ref{qm_ampl}) is that classically $\sum_{\boldsymbol k}
p_{\boldsymbol{k,j}}(t) = 1$, whereas quantum mechanically
$\sum_{\boldsymbol k} |\alpha_{\boldsymbol{k,j}}(t)|^2 =1$ holds.

For the full solution of Eqs.(\ref{mast_eq0}) and (\ref{sgl_ampl}) all the
eigenvalues {\sl and} all the eigenvectors of ${\bf T}=-{\bf H}$ (or,
equivalently, of ${\bf A}$) are needed. Let $\lambda_n$ denote the $n$th
eigenvalue of ${\bf A}$ and ${\bf \Lambda}$ the corresponding eigenvalue
matrix. Furthermore, let ${\bf Q}$ denote the matrix constructed from the
orthonormalized eigenvectors of ${\bf A}$, so that ${\bf A} = {\bf
Q}{\bf\Lambda}{\bf Q}^{-1}$. Now the classical transition probability is
given by
\be
p_{\boldsymbol{k,j}}(t) = \langle \boldsymbol k| {\bf Q} e^{-t \gamma{\bf
\Lambda}} {\bf Q}^{-1} | \boldsymbol j
\rangle,
\label{cl_prob_full}
\ee
whereas the quantum mechanical transition probability is
\be
\pi_{\boldsymbol{k,j}}(t) \equiv |\alpha_{\boldsymbol{k,j}}(t)|^2 = |\langle
\boldsymbol k| {\bf Q} e^{- i t
\gamma {\bf
\Lambda}} {\bf Q}^{-1} | \boldsymbol j \rangle|^2.
\label{qm_prob_full}
\ee

The unitarity of the time evolution operator prevents the quantum
mechanical transition probability from having a definite limit when
$t\to\infty$.  In order to compare the classical long time probability
with the quantum mechanical one, one usually uses the limiting probability
(LP), i.e., the long time average of $\pi_{\boldsymbol{k,j}}(t)$
\cite{aharonov2001}:
\be
\chi_{\boldsymbol{k,j}} \equiv \lim_{T\to\infty} \frac{1}{T} \int_0^T d t \
\pi_{\boldsymbol{k,j}}(t).
\label{limitprob}
\ee

\subsection{Boundary conditions and the Bloch ansatz}

\begin{figure}[ht]
\centerline{\includegraphics[clip=,width=0.5\columnwidth]{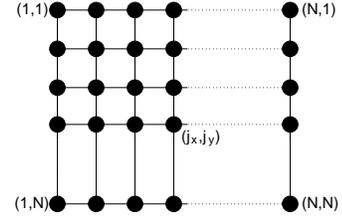}}
\caption{Sketch of a square network arranged as a regular lattice with the
appropriate numbering of the nodes. Note that the actual geometrical
realization can be much more flexible, see text for details.
}
\label{lattice}
\end{figure}

Given that our two-dimensional structures have side length $N$, they
contain $N^2$ nodes giving rise to $N^2$ basis states. We switch to a pair
notation, by which we set $|\boldsymbol j\rangle = |j_x,j_y\rangle$, where
$j_x$ and $j_y$ are integer labels in the two directions, with
$j_x,j_y\in[1,N]$, see Fig.\ref{lattice}. This labeling of the states is
not to be confused with the labeling of the adjacency matrix. Note that
capital bold letters denote matrices, while small bold letters denote the
nodes and the states.

We stop to recall a basic fact, namely that our matrix $\bf A$ does not
necessarily imply that the structure we consider obey any kind of
translational symmetry; in fact the symmetry of $\bf A$ is topological in
nature, a fact well realized both in polymer physics \cite{Gurtov2002} and
also in quantum chemistry \cite{Blumen1977}, where the $\bf A$-matrices
are fundamental in H\"uckel molecular orbital calculations
\cite{Streitwieser,McQuarrie}. Regular, translational invariant lattices
are only {\sl one} possible realization of the network structures we are
facing here. The methods of solid state physics \cite{Ziman,Kittel} apply,
however, even in our very general case.

The focus in solid state physics is on systems where Born - von Karman
periodic boundary conditions (PBC) are assumed. 
Here, all transmission rates are taken to be equal. Hence, time is
given in units of the inverse transmission rate $\gamma^{-1}$ or,
equivalently, we set $\gamma=1$. 
Now, for an internal site
of our network (not on an edge or in a corner), the Hamiltonian acting on
a state $|\boldsymbol j\rangle = |j_x,j_y\rangle$ reads
\bea
{\bf H} | j_x,j_y\rangle =&& 2 | j_x,j_y\rangle - | j_x+1,j_y\rangle - |
j_x-1,j_y\rangle \nonumber \\
&+& 2 | j_x,j_y\rangle - | j_x, j_y+1\rangle - | j_x,j_y-1\rangle.
\label{honj}
\eea
PBC extend this equation to all the sites of the network by interpreting
every integer $j_x$ and $j_y$ to be taken modulus $N$.  With this
generalization, the time independent SE
\be
{\bf H} | \Psi_{\boldsymbol\theta}\rangle = E_{\boldsymbol\theta}
|\Psi_{\boldsymbol\theta}\rangle 
\label{hbloch}
\ee
admits (as is well known) the following Bloch eigenstates
\be
|\Psi_{\boldsymbol\theta}\rangle = \frac{1}{N} \sum_{j_x,j_y=1}^{N}
e^{-i(\boldsymbol\theta \cdot \boldsymbol j)} |\boldsymbol j\rangle.
\label{blochfct}
\ee
as solutions, where $\boldsymbol\theta\cdot\boldsymbol j$ stands for the
scalar product with $\boldsymbol\theta = (\theta_x,\theta_y)$. The usual
Bloch relation can be obtained by projecting
$|\Psi_{\boldsymbol\theta}\rangle$ on the state $|\boldsymbol j\rangle$
such that $\Psi_{\boldsymbol\theta}(\boldsymbol j) \equiv \langle
\boldsymbol j | \Psi_{\boldsymbol\theta}\rangle = e^{-i(\boldsymbol\theta
\cdot \boldsymbol j)} / N$, thus $\Psi_{\boldsymbol\theta}(j_x+1,j_y+1) =
e^{-i(\theta_x + \theta_y)} \Psi_{\boldsymbol\theta}(j_x,j_y)$. The PBC
restrict the allowed values of $\boldsymbol\theta$.  In our case (side
length $N$), the PBC require that $\Psi_{\boldsymbol\theta}(N+1,j_y) =
\Psi_{\boldsymbol\theta}(1,j_y)$ and $\Psi_{\boldsymbol\theta}(j_x,N+1) =
\Psi_{\boldsymbol\theta}(j_x,1)$. It follows that one must have $\theta_x
= 2n\pi/N$ and $\theta_y=2l\pi/N$, where $n$ and $l$ are integers and
$n,l\in[1,N]$.  It is now a simple matter to verify that the
$|\Psi_{\boldsymbol\theta}\rangle$ also obey
$\langle\Psi_{\boldsymbol\theta}|\Psi_{\boldsymbol\theta'}\rangle =
\delta_{\boldsymbol{\theta,\theta'}}$ and $\sum_{\boldsymbol\theta}
|\Psi_{\boldsymbol\theta}\rangle\langle\Psi_{\boldsymbol\theta}| =
\boldsymbol 1$. Moreover, by inverting Eq.(\ref{blochfct}) one has
\be
|\boldsymbol j\rangle = \frac{1}{N} \sum_{\boldsymbol\theta} e^{i(\boldsymbol \theta
\cdot \boldsymbol j)} |\Psi_{\boldsymbol\theta}\rangle,
\ee
and $|\boldsymbol j\rangle$ might be viewed as a Wannier function of the
problem \cite{Ziman,Kittel}.

Furthermore, from Eqs.(\ref{hbloch}) and (\ref{blochfct}) the energy is
obtained as
\be
E_{\boldsymbol\theta} = 4 - 2\cos\theta_x - 2\cos\theta_y = E_{\theta_x} +
E_{\theta_y},
\label{bloch_ev}
\ee
with $E_{\theta_x} = 2 -2\cos\theta_x$ and $E_{\theta_y} = 2 -
2\cos\theta_y$. Under PBC the two-dimensional eigenvalue problem separates
into two one-dimensional problems.

The transition amplitude at time $t$ from state $|\boldsymbol j\rangle$ to
state $|\boldsymbol k\rangle$ is now, using Eq.(\ref{blochfct}) twice:
\bea
\alpha_{\boldsymbol{k,j}}(t) 
&=& 
\frac{1}{N^2} \sum_{\boldsymbol{\theta,\theta'}}
\langle\Psi_{\boldsymbol\theta'} | e^{-i(\boldsymbol\theta' \cdot
\boldsymbol k)} e^{-i{\bf
H}t} e^{i(\boldsymbol \theta \cdot \boldsymbol j)} |\Psi_{\boldsymbol \theta}\rangle 
\nonumber \\
&=& 
\frac{1}{N^2} \sum_{\boldsymbol{\theta}} e^{-i E_{\boldsymbol\theta}t} e^{-i
\boldsymbol \theta \cdot (\boldsymbol k - \boldsymbol j)}
\label{transamplbloch}
\eea
In the limit $N\to\infty$, the sums in Eq.(\ref{transamplbloch}) may be
changed to integrals; by making use of Eq.(\ref{bloch_ev}) we obtain
\bea
\lim_{N\to\infty} \alpha_{\boldsymbol{k,j}}(t) 
&=& 
\frac{e^{-i4t}}{4\pi^2}
\int\limits_{-\pi}^{\pi} d\theta_x \ e^{-i\theta_x(k_x-j_x)}
e^{i2t\cos\theta_x} 
\nonumber \\
&& \times 
\int\limits_{-\pi}^{\pi} d\theta_y \ e^{-i\theta_y(k_y-j_y)}
e^{i2t\cos\theta_y}
\\
&=& 
i^{k_x-j_x} i^{k_y-j_y} e^{-i4t} J_{k_x-j_x}(2t) J_{k_y-j_y}(2t) \nonumber ,
\eea
where $J_n(x)$ is the Bessel function of the first kind \cite{Ito}. Thus,
on a network topologically equivalent to a square lattice with PBC the
transition amplitude between the nodes $\boldsymbol j$ and $\boldsymbol k$
is given by
\be
\lim_{N\to\infty} \pi_{\boldsymbol{k,j}}(t) = [J_{k_x-j_x}(2t) J_{k_y-j_y}(2t)]^2.
\label{prob_bessel}
\ee

For systems without PBC the situation is more subtle.  If we are
interested in the behavior of a particular network of finite size $N\times
N$ we have, in general, to rely on the (numerical) solution of the
eigenvalue problem and on the calculation of the corresponding transition
probabilities. 
However, for the smallest network of $2\times 2$ nodes, which is
equivalent to a line of $4$ nodes with PBC, the analytic result for
$\pi_{\boldsymbol{k,j}}(t)$ was given in \cite{mb2005b} [see Eq.\ (19)
there]. There, the $\pi_{\boldsymbol{k,j}}(t)$ were simple trigonometric
functions. For larger networks with PBC, the $\pi_{\boldsymbol{k,j}}(t)$
are given by a multiple sum which follows from the Bloch ansatz, see
Eq.(\ref{transamplbloch}). Thus, analytic results for large finite
networks are hard to get. The situation becomes even more complex if there
are reflecting boundaries.
Evidently, the expectation is that larger systems will
still behave as in the PBC case, since for very large systems most of the
nodes are far from the boundaries. 

The discrete version of the quantum walk on one and two dimensional
networks was studied in \cite{tregenna2003}. Comparable to the periodicity
found for the CTQW on a line of $4$ nodes in \cite{mb2005b}, it was found
that in this case also the discrete quantum walks show periodic behavior,
see also \cite{travaglione2002}. In a related context, mixing properties
of {\sl discrete} quantum walks have been studied in
\cite{ahmadi2002,adamczak2003}. Here again, it was shown that a line of
$4$ nodes has a uniform mixing property, i.e.\ at certain times the
transition probabilities $\pi_{\boldsymbol{k,j}}(t)$ for all nodes
$\boldsymbol{k}$ are equal. This is also true for the CTQW on a line with
$4$ nodes \cite{mb2005b}.

\section{Probability distributions} \label{pd}

In the remainder of this paper we calculate for several finite networks of
various sizes the transition probabilities $\pi_{\boldsymbol{k,j}}$ and
their long time average $\chi_{\boldsymbol{k,j}}$. The numerical
determination of the eigenvalues and of the eigenvectors proceeded using
the FORTRAN eigensystem subroutine package (EISPACK) \cite{eispack}. 

\subsection{Eigenvalue spectra} \label{ev}

In order to have a comparison to our calculations, we contrast the
numerically obtained eigenvalue spectra for finite networks to the results
obtained for PBC from the Bloch ansatz, Eq.(\ref{bloch_ev}). We note from
the start that Eq.(\ref{bloch_ev}) limits the latter eigenvalues to the
interval $[0,8]$. Furthermore, the eigenvalues $0$ and $8$ are
nondegenerate because they are only obtained for $n=l=N$ and for
$n=l=N/2$, respectively. 

\begin{figure}[ht]
\centerline{\includegraphics[clip=,width=0.9\columnwidth]{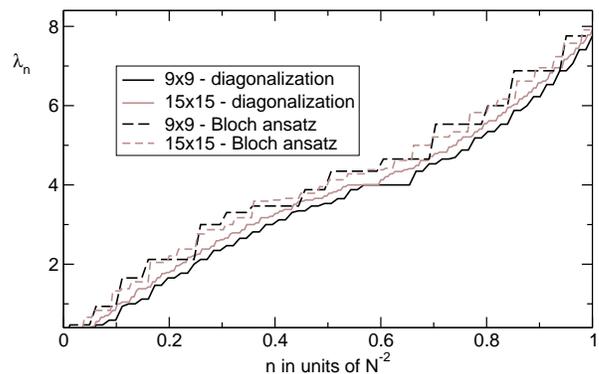}}
\caption{Eigenvalues $\lambda_n$, arranged in ascending order, for the
networks of size $N=9$ (black) and $N=15$ (grey) obtained for the finite
network, solid lines, and from the Bloch solutions with PBC, dashed lines.
}
\label{ev_latt}
\end{figure}

Figure \ref{ev_latt} shows the eigenvalue spectra for two networks with
sizes $N=9$ and $N=15$ each. Displayed are the results for the finite
networks obtained by numerical diagonalization and for the PBC cases from
the Bloch solutions. 
The numerical results for PBC agree exactly with the Bloch solutions
within the precision of our calculations and would be indistinguishable
from each other in Fig.\ref{ev_latt}, therefore we display only the
former.
We note that the numerically determined spectra are
also bound by $0$ and by $8$, although the value $8$ is only approached in
the limit $N\to \infty$. 

\begin{figure}[ht]
\centerline{\includegraphics[clip=,width=0.9\columnwidth]{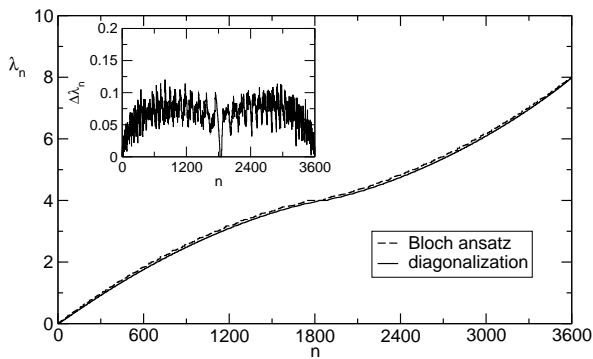}}
\caption{Eigenvalues $\lambda_n$, arranged in ascending order, for the
network of size $N=60$. The numerical result for the finite network, solid
line, is compared to the Bloch solution with PBC, dashed line. The inset
shows the difference between these two results as a function of $n$.
}
\label{ev_latt60}
\end{figure}

Figure \ref{ev_latt60} displays the findings for a larger network, $N=60$,
again comparing the spectra of the finite network with those obtained from
Eq.(\ref{bloch_ev}).  We again note that the spectra get to be very close
(a fact which was to be expected because now $N$ is larger than in
Fig.\ref{ev_latt}), but that the PBC case still lies systematically
higher. To highlight this fact, we plotted in the inset of
Fig.\ref{ev_latt60} for each $n$ the difference in the values between the
$\lambda_n$ obtained with and without PBC.

\subsection{Special initial conditions}

We now turn to the dynamics of the propagation through the network and
focus on special initial points. In doing so, we continue to compare our
results for the finite network to the Bloch solutions. The computations
for the finite network without PBC are performed by calculating all
eigenvalues $\lambda_n$ and all eigenvectors of the matrix $\bf A$ using
the FORTRAN EISPACK routine. The Bloch solutions are obtained by using the
standard software packages MAPLE 7. 

\begin{figure}[ht]
\centerline{\includegraphics[clip=,width=0.95\columnwidth]{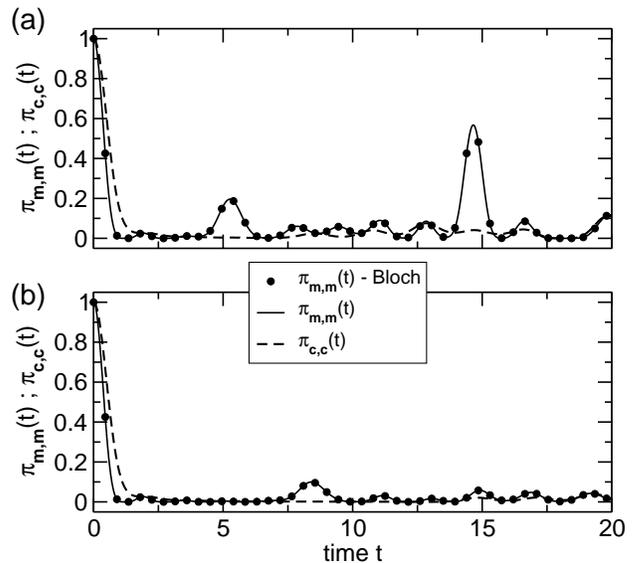}}
\caption{Probabilities for a CTQW to be at the initial site at time $t$ on
networks of size (a) $N=9$ and (b) $N=15$. The starting sites are the
middle node $\boldsymbol m$ (solid lines) and the corner node $\boldsymbol
c$ (dashed lines), respectively.  The results for the finite network,
obtained by using Eq.(\ref{qm_prob_full}), solid and dashed lines, are
compared to the Bloch solutions with PBC, obtained via
Eq.(\ref{transamplbloch}), dots, see text for details.  Time is given in
units of the inverse transmission rate $\gamma^{-1}$.  
}
\label{cmp_bloch}
\end{figure}

We begin by choosing the middle node $\boldsymbol m$ of the network as the
starting site.  Figure \ref{cmp_bloch} shows the probability of being at
$\boldsymbol m$ after time $t$ obtained via Eq.(\ref{qm_prob_full}) for
two network sizes, $N=9$ and $N=15$, respectively. Moreover, we have also
inserted into Fig.\ref{cmp_bloch} the results obtained by starting at a
corner node $\boldsymbol c$. Clearly, on short time scales our numerical
results for $\pi_{\boldsymbol{m,m}}(t)$, solid line, agree nicely with the
Bloch solutions, dots, given in Eq.(\ref{transamplbloch}), even for the
$N=9$ network. However, the probability $\pi_{\boldsymbol{cc}}(t)$ of
being at the initial corner node $\boldsymbol c$ after time $t$, dashed
line, differs quite early from the Bloch solution, given that now the
boundaries play an important role. 

\begin{figure}[h]
\centerline{\includegraphics[clip,width=0.95\columnwidth]{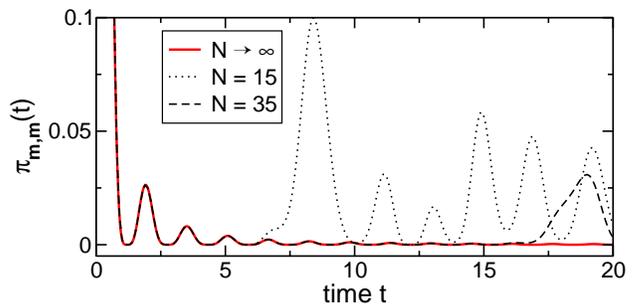}}
\caption{Comparison of the probabilities $\pi_{\boldsymbol{m,m}}(t)$ for
network sizes $N=15$ and $N=35$ with the bulk system where $N\to\infty$.
Time is given in units of
$\gamma^{-1}$.}
\label{cmp_pqm_bloch}
\end{figure}

In order to show for finite networks the deviations from the Bloch
solutions, we plot in Fig.\ref{cmp_pqm_bloch}, for networks of sizes
$N=15$ and $N=35$, the probabilities $\pi_{\boldsymbol{m,m}}(t)$ to be at
time $t$ at the middle node $\boldsymbol{m}$, while also starting at
$\boldsymbol{m}$.  On short time scales, the $\pi_{\boldsymbol{m,m}}(t)$
behave exactly as the Bloch solution for the bulk system given in
Eq.(\ref{prob_bessel}).  However, for finite networks at longer times
there is constructive interference due to the reflections at the
boundaries, which result in a higher probability to find the walker back
at $\boldsymbol{m}$. This interference, of course, can only take place
after the wave has crossed the whole network and returns to
$\boldsymbol{m}$. For the two examples given in Fig.\ref{cmp_pqm_bloch},
this happens approximately at the times $\gamma t\approx6.1$ for $N=15$
and $\gamma t\approx15.6$ for $N=35$. As for the one dimensional case
studied in \cite{mb2005b}, such deviations are found in general at times
around $N/2$.

\begin{figure}[ht]
\centerline{\includegraphics[clip=,width=0.95\columnwidth]{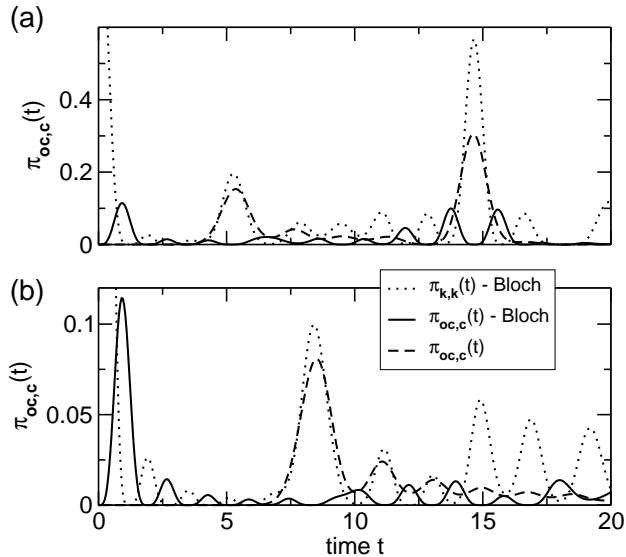}}
\caption{Probabilities for a CTQW starting at one corner node $\boldsymbol
c$ to be at the opposite corner node $\boldsymbol{oc}$ at time $t$ on
networks of size (a) $N=9$ and (b) $N=15$.  The results for the finite
network, obtained by using Eq.(\ref{qm_prob_full}), dashed lines, are
compared to the Bloch solutions with PBC, obtained via
Eq.(\ref{transamplbloch}), black solid lines. We further compare also to
the Bloch solution with PBC, dotted lines, where the initial and final node
is the node $\boldsymbol k$, see Fig.\ref{cmp_bloch} and text for details.
Time is given in units of the inverse transmission rate $\gamma^{-1}$.  
}
\label{cornertime}
\end{figure}

In Fig.\ref{cornertime} we show the transition probabilities
$\pi_{\boldsymbol{oc,c}}(t)$ to go from one corner node $\boldsymbol c$ to
the opposite corner node $\boldsymbol{oc}$ in time $t$. Again we take
$N=9$ and $N=15$ as network sizes. We remark that already after a short
period of time there is a considerable probability for the CTQW to be at
the opposite node. For example, for the $N=9$ network we find that
$\pi_{\boldsymbol{oc},\boldsymbol c}(t) \approx 0.16$ at $\gamma t\approx
5.4$. This means that for $\gamma t\approx 5.4$ a remaining probability of
$0.84$ is distributed among the other $80$ nodes, which is roughly an
order of magnitude less than $0.16$.

Not only is there a very high probability to go to the opposite node, but
also is the transport to this node very fast. The same holds for the
$N=15$ network. Here, the first peak of $\pi_{\boldsymbol{oc},\boldsymbol
c}(t)$ occurs at about $\gamma t\approx 8.4$, for which
$\pi_{\boldsymbol{oc},\boldsymbol c}(t) \approx 0.08$, which is again a
relatively high value. We also remark that the shortest ``chemical''
distance between two opposite corner nodes on this network is $\Delta x =
28$ bonds. Therefore, the (initial) ``velocity'' of the CTQW is $\Delta x
/ \gamma t \approx 3.3$.

For a closer examination, we also confront our calculations to the Bloch
solution, see Fig.\ref{cornertime}. The results from the Bloch ansatz
naturally compare only to ours where the middle node is the initial
node. However, we note that the maxima of the probability
$\pi_{\boldsymbol{oc,c}}(t)$ on the finite networks without PBC from a
corner $\boldsymbol c$ to the opposite corner $\boldsymbol{oc}$ occur at
approximately the same positions as the ones for
$\pi_{\boldsymbol{k,k}}(t)$ from the Bloch solution for CTQWs with PBC
from any node $\boldsymbol k$ to the same node in the same time $t$. This
is quite remarkable because the distances traveled by the CTQWs are
different in both cases. For the finite network of size $N=15$ without
PBC, there are $28$ bonds from $\boldsymbol c$ to $\boldsymbol{oc}$,
whereas for a network with PBC there are always $15$ bonds from
$\boldsymbol k$ to a nearest site corresponding to the same $\boldsymbol
k$; for the $N=9$ network, the corresponding distances are $16$ bonds and
$9$ bonds, respectively. This implies that in this particular situation
the initial ``velocity'' for the finite network is higher than the one for
the network with PBC.

\begin{figure}[ht]
\centerline{\includegraphics[clip=,width=0.95\columnwidth]{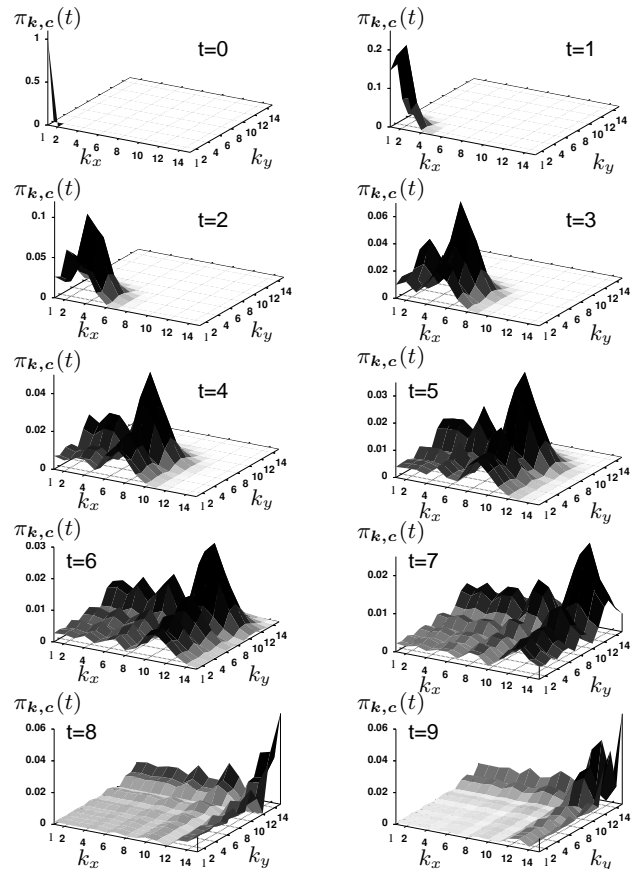}}
\caption{Snapshots of the probabilities $\pi_{\boldsymbol{k,c}}$ to be at
time $t$ at node $\boldsymbol k$ when starting at node $\boldsymbol c
=(1,1)$. Here the network is of size $N=15$ and time is given in units of
$\gamma^{-1}$.  
}
\label{3dcornertime}
\end{figure}

In Fig.\ref{3dcornertime} we display snapshots in time of the transition
probabilities $\pi_{\boldsymbol k,\boldsymbol c}(t)$ to go from the corner
node $\boldsymbol c = (1,1)$ to the other nodes.  We see again that on
short time scales the transport is very fast and, furthermore, that the
main fraction of the probability stays on the diagonal. One might think
that the effect is ballistic, but this is not the case: neighboring sites
along the diagonal are not directly related via $\bf A$ (one may also
remember that in $\bf A$ only the {\sl topology} matters). The observed
effect is quantum mechanical in nature: on the diagonal sites constructive
interferences are particularly manifest.

The phenomenon of fast CTQW transport through the network can also be
observed when starting at any other node, where on short time scales we
find a high probability of going to the ``mirror'' node. We define the
``mirror'' node of $(j_x,j_y)$ to be $(N+1-j_x,N+1-j_y)$, so
that the two nodes are related by inversion with respect to the center of
the network. 

This selective behavior has to be contrasted with the one displayed by
classical CTRWs. These describe namely simple diffusion, in which no
exceptional transition probabilities between particular nodes occurs. 

\subsection{Limiting probabilities}

We continue by examining the situation at even longer times and focus in
particular on the LPs $\chi_{\boldsymbol{k,j}}$ given by
Eq.(\ref{limitprob}). Here, because we had already determined all
eigenvalues and eigenvectors, we took advantage of the structure of
Eq.(\ref{limitprob}). From Eqs.(\ref{qm_ampl}) and (\ref{limitprob}), and
denoting the orthonormalized eigenstates of the Hamiltonian by
$|\boldsymbol q_n\rangle$, such that $\sum_n |\boldsymbol q_n\rangle
\langle \boldsymbol q_n | = \boldsymbol 1$, we find that
\begin{subequations}
\begin{align}
\chi_{\boldsymbol{k,j}} &=
\lim_{T\to\infty}\frac{1}{T} \int\limits_0^T dt \ \left|
\sum_{n}\langle \boldsymbol k | e^{-i{\bf H}t} |
\boldsymbol q_n \rangle \langle \boldsymbol q_n |
\boldsymbol j \rangle \right|^2 
\nonumber \\
&=
\lim_{T\to\infty}\frac{1}{T} \int\limits_0^T dt \ \left|
\sum_{n}e^{-i\gamma\lambda_n t} \langle \boldsymbol k
| \boldsymbol q_n \rangle \langle \boldsymbol q_n |
\boldsymbol j \rangle \right|^2 
\nonumber \\
&=
\sum_{n,m} \langle \boldsymbol k |
\boldsymbol q_n \rangle \langle \boldsymbol q_n |
\boldsymbol j \rangle \langle \boldsymbol j | \boldsymbol q_m
\rangle \langle \boldsymbol q_m | \boldsymbol k \rangle
\nonumber \\
&
\times 
\left( \lim_{T\to\infty}\frac{1}{T} \int\limits_0^T dt \
e^{-i(\lambda_n - \lambda_m)\gamma t} \right)
\label{limprob_ev_int}
\\
&=
\sum_{n,m} 
\delta_{\lambda_n,\lambda_m}
\langle \boldsymbol k |
\boldsymbol q_n \rangle \langle \boldsymbol q_n |
\boldsymbol j \rangle \langle \boldsymbol j | \boldsymbol q_m
\rangle \langle \boldsymbol q_m | \boldsymbol k \rangle.
\label{limprob_ev}
\end{align}
\end{subequations}
We note that the integral in Eq.(\ref{limprob_ev_int}) equals $1$ if
$\lambda_n = \lambda_m$ and $0$ otherwise, i.e., it equals
$\delta_{\lambda_n,\lambda_m}$.  Given that some eigenvalues of $\bf H$
are degenerate, the sum in Eq.(\ref{limprob_ev}) can contain terms
belonging to different eigenstates $|\boldsymbol q_n\rangle$ and
$|\boldsymbol q_m\rangle$. Equation (\ref{limprob_ev}) provides a
numerically very efficient way of computing the $\chi_{\boldsymbol{k,j}}$.
Remarkably, we find that the $\chi_{\boldsymbol{k,j}}$ depend in an
unexpected way on the exact value of the size $N$ of the finite network
under study.  Given these unexpected findings, which we report in the
following, we cross-checked our evaluation method based on
Eq.(\ref{limprob_ev}) very carefully, by comparing it in selected cases to
the direct evaluation of the integral in Eq.(\ref{limitprob}). In so
doing, we fixed the upper integration limit to a very large value and
verified that even larger values didn't lead to any changes in
$\chi_{\boldsymbol{k,j}}$. In all cases we found that both numerical
methods agree to very high precision. Thus, in general we prefered to work
with Eq.(\ref{limprob_ev}), which is computationally much faster than
Eq.(\ref{limitprob}).

\begin{figure}[h]
\centerline{\includegraphics[clip=,width=0.95\columnwidth]{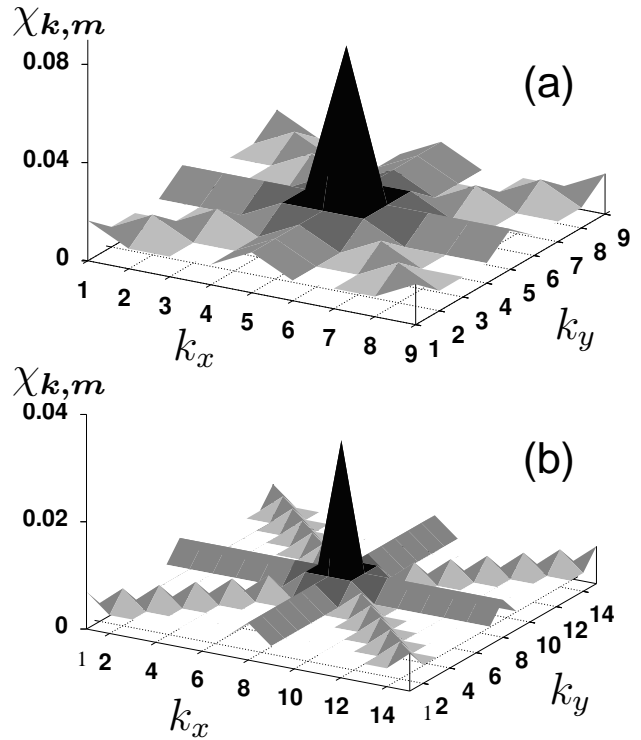}}
\caption{Limiting probabilities $\chi_{\boldsymbol{k,m}}$ to be at node
$\boldsymbol k$ when starting at the middle node $\boldsymbol m$ for
networks of sizes (a) $N=9$ and (b) $N=15$. The results are obtained from
Eq.(\ref{limprob_ev}).
}
\label{limprobcomp}
\end{figure}

In Fig.\ref{limprobcomp} we present $\chi_{\boldsymbol{k,m}}$ for CTQWs,
in which the starting node is the middle node $\boldsymbol m$. We display
results for networks of sizes $N=9$ and $N=15$. Note that the
$\chi_{\boldsymbol{k,m}}$ are symmetric about the initial middle node
$\boldsymbol m$, meaning that a node and its ``mirror'' node have the same
limiting probabilities. This is in no way surprising, because this
symmetry is already inherent in $\bf A$ and $\bf H$. More remarkable are
the patterns obtained. They may be contrasted to the classical CTRW, in
which the limiting probability distribution is uniform, thus symmetric for
all nodes.

~\\

Also when starting at a corner node $\boldsymbol c$, we often find that
the LPs for the starting node and its ``mirror'' node are equal.  In
Fig.\ref{limprobsym} we show the $\chi_{\boldsymbol{k,c}}$ obtained by
going from the corner node $\boldsymbol c=(1,1)$ to the other nodes for
networks of sizes $N=5$, $N=14$, $N=23$, and $N=47$. 

However, for some particular network sizes the distributions of the LPs
turn out to be asymmetric. For instance, for a network of size $N=15$ the
LP $\chi_{\boldsymbol{oc,c}}$ for the CTQW starting at node $\boldsymbol
c$ to be at the opposite corner node $\boldsymbol{oc}$ is less than the LP
$\chi_{\boldsymbol{c,c}}$ to be at the initial node. The same is true for
the nodes along the edges of the network.  Figure \ref{limprobantisym}
shows that such asymmetries occur for networks of the sizes $N=6$, $N=15$,
$N=24$, and $N=48$ (the asymmetries are best seen by looking at
$\chi_{\boldsymbol{c,c}}$ and $\chi_{\boldsymbol{oc,c}}$). Note that these
asymmetries  occur for networks in which $N$ is increased only by unity
compared to networks which behave symmetrically, e.g., see
Fig.\ref{limprobsym}.  The smallest network where we detected asymmetries
in the distribution of the LPs has $N=6$. The next ones we found for
$N=12, 15, 18, 21, 24, 30, 36, \cdots$.  

\begin{widetext}
~
\begin{figure}[ht]
\centerline{\includegraphics[clip=,width=0.85\columnwidth]{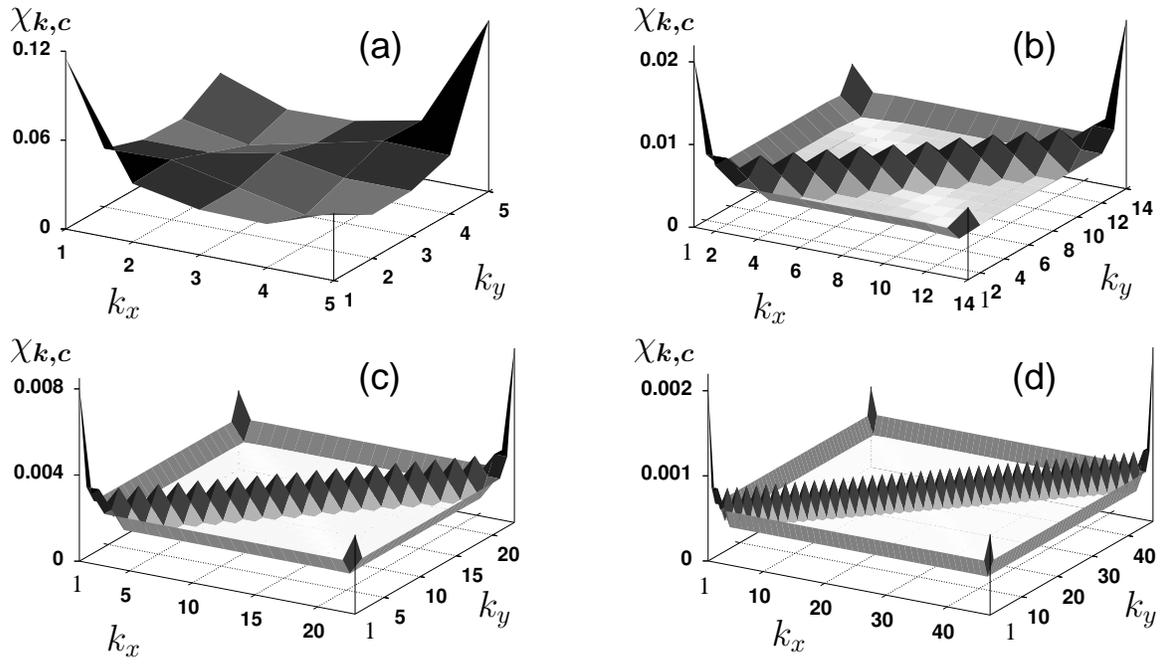}}
\caption{As in Fig.\ref{limprobcomp}, LPs $\chi_{\boldsymbol{k,c}}$ to be
at node $\boldsymbol k$ when starting at the corner node $\boldsymbol c =
(1,1)$ for networks of sizes (a) $N=5$, (b) $N=14$, (c) $N=23$, and (d)
$N=47$.
}
\label{limprobsym}
\end{figure}

~\\~\\~\\~\\

\begin{figure}[ht]
\centerline{\includegraphics[clip=,width=0.85\columnwidth]{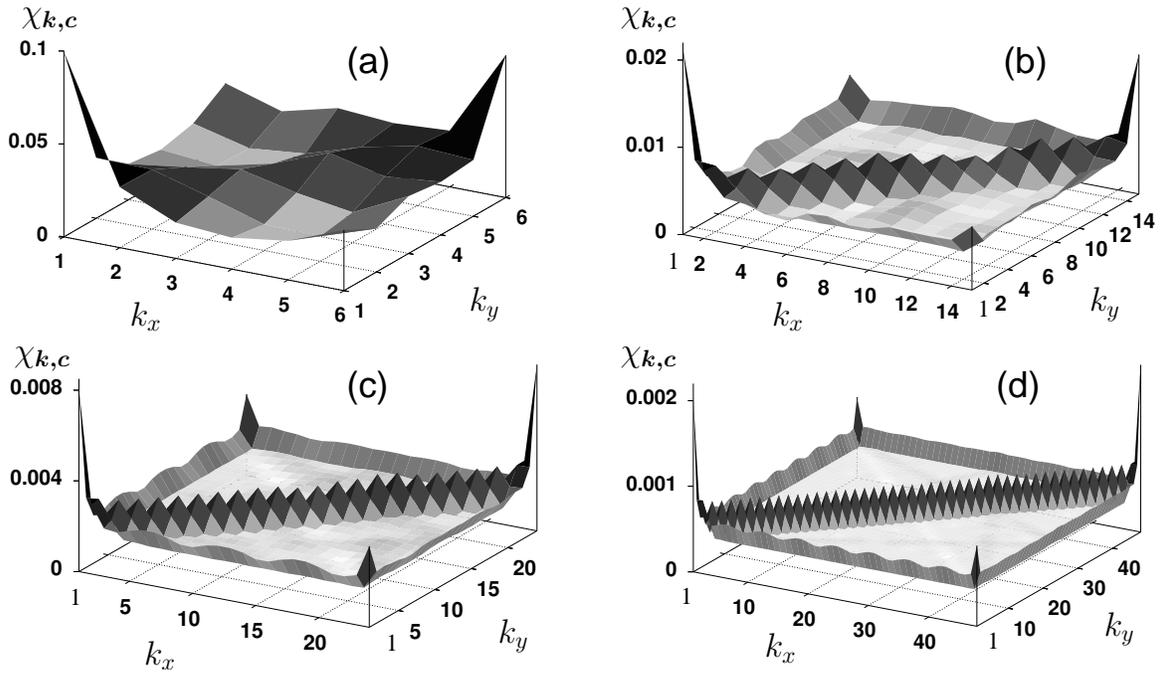}}
\caption{As in Fig.\ref{limprobcomp}, LPs $\chi_{\boldsymbol{k,c}}$ to be
at node $\boldsymbol k$ when starting at the corner node $\boldsymbol c =
(1,1)$ for networks of sizes (a) $N=6$, (b) $N=15$, (c) $N=24$, and (d)
$N=48$.  One may note the asymmetries by comparing to
Fig.\ref{limprobsym}.
}
\label{limprobantisym}
\end{figure}
\end{widetext}

\begin{figure}[h]
\centerline{\includegraphics[clip=,width=0.95\columnwidth]{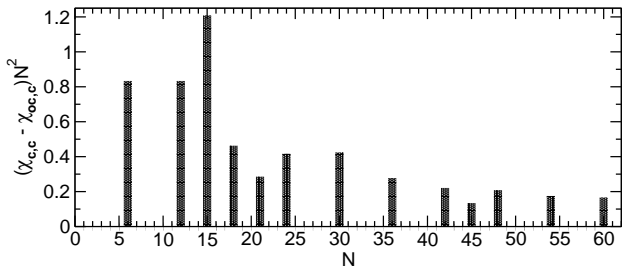}}
\caption{Differences between the LPs for CTQWs that start at $\boldsymbol
c=(1,1)$ to be at $\boldsymbol c$, $\chi_{\boldsymbol c,\boldsymbol c}$,
or to be at its ``mirror'' node $\boldsymbol{oc}=(N,N)$,
$\chi_{\boldsymbol{oc},\boldsymbol c}$, as a function of the network size
$N$, for $1\leq N \leq60$.
}
\label{limprobdiff}
\end{figure}

\begin{figure}[h]
\centerline{\includegraphics[clip=,width=0.95\columnwidth]{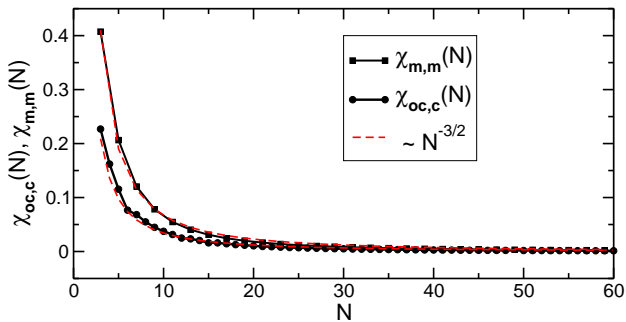}}
\caption{$N$-dependence of the LPs $\chi_{\boldsymbol{oc,c}}(N)$ and
$\chi_{\boldsymbol{m,m}}(N)$. For $\chi_{\boldsymbol{m,m}}(N)$ only odd
numbered networks are shown, since only those have a single central node
$\boldsymbol{m}$.
}
\label{limprobN}
\end{figure}

The asymmetries are small and therefore not easy to spot in the global
pictures displayed in Figs.\ref{limprobsym} and \ref{limprobantisym}. As
illustrative examples, we choose prominent points in the network to show
the asymmetries.
An asymmetric LP distribution is particularly evident in the difference
between $\chi_{\boldsymbol{c,c}}$ and $\chi_{\boldsymbol{oc,c}}$.  Thus,
as an overview we present in Fig.\ref{limprobdiff} as a function of $N$ a
plot of the $(\chi_{\boldsymbol{c,c}} - \chi_{\boldsymbol{oc,c}}) N^2$
values obtained. Note that all $N$ values in Fig.\ref{limprobdiff} for
which $(\chi_{\boldsymbol{c,c}} - \chi_{\boldsymbol{oc,c}}) \neq 0$ are
divisible by $3$.  However, the converse is not true, we find symmetric LP
distributions for the networks with $N=3, 9, 27, 33, 39, \cdots$. 
The general $N$-dependence of $\chi_{\boldsymbol{oc,c}}(N)$ and
$\chi_{\boldsymbol{m,m}}(N)$ is plotted in Fig.\ref{limprobN}. In contrast
to the classical limiting probability, which in all cases shows
equipartition between all nodes, i.e.\ it is given by $N^{-2}$, we find
that $\chi_{\boldsymbol{oc,c}}(N)$ and $\chi_{\boldsymbol{m,m}}(N)$ decay
nearly algebraically, namely as $N^{-3/2}$.

Another striking feature of the CTQW is that the LPs display quite regular
patterns over the network. For the CTQW starting at the network's middle
node $\boldsymbol m$, the distributions of the LPs show a star-like
pattern, see Fig.\ref{limprobcomp}. In all cases studied here, the LP
distribution is such that its major fraction appears to be distributed
along lines diagonal and parallel to the network's edges and crossing each
other at the initial node. Here again this implies that the transport is
generated by constructive and destructive interference. This might have
consequences: On a regular, square network (or on a lattice for
that matter) the application of the CTQW as a search algorithm is flawed,
since its effectiveness is correlated with the initial site. That is,
there is a high probability of finding a certain node which is
``constructively'' correlated with the initial one, but there is also a
rather low probability of finding the others. Thus, although the topology
of the square network has no exceptional sites, the transport through the
lattice strongly depends on the initial condition. This relates directly
to previous studies, where it was found that quantum transport can become
much slower than the classical one \cite{mb2005a}.
However, there it was also shown that if one starts in a superposition of
states, the transport can get to be much quicker than in the classical case. A
similar effect should also be observable here. For instance, if one starts
in a uniform superposition of states along the baseline of the network,
the CTQW can be mapped onto a one dimensional problem, as treated for
instance in \cite{childs2002}.

Furthermore, the fact that the LP distributions have especially high peaks
at the initial node and at its ``mirror'' node strongly recalls (in the
spatially discrete version discussed here) the quantum mirage effects
found in elliptic quantum corrals; these, again, can be related to wave
interferences, see for instance \cite{manoharan2000,fiete2003}.  A more
detailed study of this effect will be published elsewhere \cite{mvb2005b}.

\section{Conclusion} \label{con}

We have studied numerically continuous-time quantum walks on finite
networks topologically equivalent to square lattices. Furthermore, we
compared our results to analytic expressions obtained from the Bloch
ansatz for networks with periodic boundary conditions. For these quantum
walks, we have found that on short time scales a directed transport
through the network takes place. In particular, when placing the initial
excitation at one corner node, the walks propagate in a rather direct
fashion along the diagonal to the opposite corner node. The transport is
not ballistic, but is rather due to constructive quantum mechanical
interferences.

In the long time limit, we found that walks on networks of specific sizes
$N$ may show (in a totally unexpected way) asymmetric limiting probability
distributions. This asymmetry manifests itself in the fact that the
limiting probabilities for a CTQW to be at the initial node $\boldsymbol c
= (1,1)$ and at its ``mirror'' node $\boldsymbol{oc} = (N,N)$ differ.
However, we were unable to find a way to predict which particular $N$
values are related to such asymmetries. 

In general, the limiting probability distributions show patterns which
depend on the starting site of the CTQW. This is a remarkable effect,
which might have 
consequences for search algorithms based on CTQWs.
Furthermore, we also found in all our calculations that the limiting
probability distributions show strong peaks at the initial node and its
``mirror'' node.  This effect resembles a discrete version of quantum
mirages.

\section*{Acknowledgments}

This work was supported by a grant from the Ministry of Science, Research
and the Arts of Baden-W\"urttemberg (AZ: 24-7532.23-11-11/1). Further
support from the Deutsche Forschungsgemeinschaft (DFG) and the Fonds der
Chemischen Industrie is gratefully acknowledged. O.\ M.\ thanks Martin
Holthaus for very helpful discussions on the topic.

\end{document}